\documentclass[useAMS,usenatbib]{mn2e}
\usepackage{psfig}
\usepackage{amsmath}
\usepackage{amssymb}
\usepackage{graphicx}
\usepackage{color}
\usepackage{hyperref}
\hypersetup{
    unicode=true,                     % non-Latin characters in Acrobat's bookmarks
    linktocpage=true,
    pdftoolbar=true,                % show Acrobat's toolbar?
   pdfmenubar=true,              % show Acrobat's menu?
    pdffitwindow=true,           % window fit to page when opened
    pdfstartview={FitH},        % fits the width of the page to the window
    pdfhighlight={/I},
   hyperindex=true,
    pdftitle={My title},            % title
   pdfauthor={Author},         % author
   pdfsubject={Subject},       % subject of the document
   pdfcreator={Creator},       % creator of the document
    pdfproducer={Producer}, % producer of the document
    pdfkeywords={keywords}, % list of keywords
    pdfnewwindow=true,      % links in new window
   colorlinks=true,       % false: boxed links; true: colored links
    linkcolor=red,          % color of internal links
    citecolor=blue,        % color of links to bibliography
    filecolor=magenta,      % color of file links
    urlcolor=cyan           % color of external links
}

\def \rqr {$r_{0QR}$}

\newcommand{\tcol}{\textcolor{black}}

\title[The mJy radio luminosity function]{Counting quasar--radio source pairs to derive the millijansky radio luminosity function and clustering strength to $z=3.5$}

\author[Fine et al.]
       {S. Fine$^1$\thanks{stephen.fine@durham.ac.uk},
         T. Shanks$^2$, R. Johnston$^1$,  M.~J. Jarvis$^{1,3}$, T. Mauch$^4$ \\
$^1$Department of Physics, University of Western Cape, Bellville 7535, Cape Town, South Africa\\
$^2$Department of Physics, Durham University, South Road, Durham DH1
         3LE, UK \\
$^3$Oxford Astrophysics, Denys Wilkinson Building, Keble Rd, Oxford
       OX1 3RH, UK \\
$^4$SKA Africa, 3rd Floor, The Park, Park Road, Pinelands 7405, South Africa  \\
}

\begin{document}

\maketitle

\begin{abstract}

\tcol{We apply a cross-correlation technique to  infer the $S>3$mJy radio luminosity function (RLF) from the NRAO VLA sky survey (NVSS) to $z\sim3.5$. We measure $\Sigma$ the over density of radio sources around spectroscopically confirmed quasars. $\Sigma$ is related to the space density of radio sources at the distance of the quasars and the clustering strength between the two samples, hence knowledge of one constrains the other. Under simple assumptions we find $\Phi\propto (1+z)^{3.7\pm0.7}$ out to $z\sim2$. Above this redshift the evolution slows and we constrain the evolution exponent to $<1.01$ ($2\sigma$). This behaviour is almost identical to that found by previous authors for the bright end of the RLF potentially indicating that we are looking at the same population. This suggests that the NVSS is dominated by a single population; most likely radio sources associated with high-excitation cold-mode accretion.  Inversely, by adopting a previously modelled  RLF we can  constrain the clustering of high-redshift radio sources and find a clustering strength consistent with $r_0=15.0\pm 2.5$\,Mpc up to $z\sim3.5$. This is inconsistent with quasars at low redshift and some measurements of the clustering of bright FRII sources. This behaviour is more consistent with the clustering of lower luminosity radio galaxies in the local universe. Our results indicate that the high-excitation systems dominating our sample are hosted in the most massive galaxies at all redshifts sampled.}

\end{abstract}


\section{Introduction}

The standard approach to measuring the radio luminosity function (RLF) requires a sample with distance information to convert fluxes to luminosities. These distances typically come from a cross-match to existing optical redshift surveys. Millijansky radio sources have sky densities of a few 10s per square degree \citep{con98,mau03} hence require wide-field ($\gtrsim100$\,deg$^2$) spectroscopic surveys to build up significant statistics in the RLF. These samples exist in the local Universe (e.g. 6dF, SDSS) and in combination with wide-field mJy radio catalogues the local RLF has been shown to be a combination of two main populations: active galactic nuclei (AGN) with a double power law LF (similar to that of quasars e.g. \citealt{boy00}) at high luminosities and star-forming galaxies with a Schechter function LF at lower luminosities \citep{bes05,mau07}.

Wide-field spectroscopic surveys are not deep enough to probe the overall galaxy population at higher redshift. Luminous red galaxies (LRGs) are bright enough to produce large samples up to $z\sim0.7$ \citep{eis01,can06}. While this is hardly a representative slice of the galaxy population, local surveys show that most radio sources in the $10^{24}<L<10^{26}$\,W/Hz regime (that translates to fluxes of $S\sim1$ to 100\,mJy at $z\sim0.7$) are associated with massive red galaxies \citep{c+b88,mau07}. Using LRGs, \citet{sad07} found evolution described well by shifting the AGN portion of the local RLF in the luminosity direction by $(1+z)^{2.0}$. These results were in broad agreement with \citet{c+j04} who used galaxies from the Sloan digital sky survey to show that fainter $L<10^{25}$\,W\,Hz$^{-1}$\,sr$^{-1}$ radio sources evolved more slowly than brighter ones up to $z\sim0.5$

Deep pencil-beam optical surveys offer higher-redshift galaxy samples that, when combined with deep radio imaging, constrain the sub-mJy RLF. At these lower flux densities ($\lesssim0.1$\,mJy) radio surveys become dominated by star forming galaxies \citep{sey08,pad09} that show strong luminosity evolution $\propto (1+z)^{\sim2.5}$ \citep{pad11,mca13}, with some contribution from radio-quiet AGN \citep{j+r04,sim06}. The lower-luminosity AGN found in these surveys show somewhat less evolution than the \citet{sad07} result. \citet{pad11} find no evolution in their AGN to $z\sim5$, and when they remove possible star-formation derived emission they find negative evolution. They suggest this may be a result of extremely high redshifts objects in their sample and the RLF cutting off and declining for $z\gtrsim 1-2$. Below these extreme redshifts \citet{smo09,mca13} find slow but significant evolution in their AGN: $\propto (1+z)^{1.2}$ and $(1+z)^{0.8}$ respectively.

Small radio samples with complete spectroscopic coverage constrain the bright end of the RLF at high $z$ \citep{d+p90,wil01}. These studies show that at bright fluxes radio sources are found up to high redshift ($z\sim3$), indicating the difficulty in obtaining complete spectroscopy on large radio samples. They also found strong evolution in the RLF.
\citet{wil01} used a combination of tiered radio samples with a faintest limit of \tcol{$S_{151MHz}>500$\,mJy to model the RLF.}
They separated their LF model into two populations roughly separated by being above or below $L\sim 10^{26}$\,W/Hz. The lower luminosity population being primarily FRI objects or FRIIs that show little evidence for an AGN in the optical, and the higher luminosity sample containing bright FRII sources often associated with optical quasars. The brighter population's LF increases towards higher $z$ peaking at $z\sim 2$ and then falling. The fainter end is described by a Schechter function that increases until it reaches $z\sim1$ after which it remains stationary. In reality the lower luminosity population is poorly constrained for $z\gtrsim 1$ although further strong evolution is ruled out by source counts.

Above $z\sim 0.7$ the mJy RLF is difficult to constrain. It lies between the parameter spaces constrained by pencil-beam surveys that run out of radio sources at higher flux densities, and the targeted surveys that require large amounts of telescope time to push fainter. In this regime the RLF has been estimated from samples that have semi-complete spectroscopic coverage supplemented by photometric redshifts. \citet{wad01} used a 1\,mJy limited sample of 72 galaxies with $65$\%\ spectroscopic completeness to show that the evolution of fainter $\sim10^{24}$\,W radio sources peaks later compared with brighter $\sim10^{26}$ sources. \citet{rig11} used a tiered sample that included the \citet{wad01} sample and pushed further down to 0.1\,mJy at 1.4\,GHz (for $z<1.3$) using photometric redshifts from the COSMOS field. They confirmed this differential evolution with radio luminosity analogous to `downsizing' seen in star formation rates and Xray/optical AGN.

The flux range $1\lesssim S \lesssim 100$\,mJy is of particular interest since it samples the RLF in the luminosity regime where the bulk of the energy density from AGN is emitted from redshifts $0.5\lesssim z \lesssim 3.5$; the peak of AGN activity in the Universe. Hence this flux range is fundamental to our understanding of radio AGN and their impact on their surrounds. Constraining the LF in this parameter space is difficult and has thus far only been possible in small samples with incomplete spectroscopy. In this paper we look at an alternative approach. We use spectroscopic quasars as a tracer of the large-scale structure at high redshift and cross correlate these with the NVSS to determine the RLF.

Throughout this work we will assume a standard flat $(\Omega_{\rm m},\Omega_{\Lambda})=(0.3,0.7)$, $h=0.7$ cosmology. The paper is organised such that we give the background to our technique in section~\ref{sec:oot}, in section~\ref{sec:data} we introduce the data sets we will be using for our analysis that is described in section~\ref{sec:anal}. In section~\ref{sec:results} we show our results and discuss their meaning for the RLF in section~\ref{sec:rlf} and high-redshift clustering in section~\ref{sec:clust}. We summarise our results in section~\ref{sec:sum}.

\section{Review of technique}
\label{sec:oot}

\tcol{The technique we follow exploits a data set that has redshift information to constrain a sample that does not  \citep[][]{phi85,p+s87}.  This approach has been used for many years and  has been recently exploited to reproduce the redshift distribution of photometric samples \citep{new08,mat10} and similarly to calibrate photometric redshifts \citep{sch10a}. Here we briefly outline the process we will follow as described in \citet{phi85} and \citet{p+s87}.}

We begin with the real-space correlation function $\xi(r,z)$ defined such that, for a galaxy population with space density $\phi(L,z)$, the probability of finding a galaxy in a volume $\delta V$ a distance $r$ from an arbitrary galaxy is
\begin{equation}
\delta P = \phi(L,z)[1+\xi(L,z,r)]\delta V.
\end{equation}
In the linear haloe-haloe regime ($1\lesssim r \lesssim 100$\,Mpc) the correlation function is well-described by a power law $\xi = (r_0/r)^\gamma$ with $\gamma\sim 1.8$ (e.g. \citealt{pee80}).

The angular statistic $\Sigma_{excess}$ is defined as the excess number of galaxies with luminosity $L$ to $L+\delta L$ within a projected radius $R$ of an arbitrary galaxy with known redshift $z$. Assuming a power law form for the correlation function, and that the evolution of $r_0$ and $\phi$ are minimal over a clustering length
\begin{equation}
\Sigma_{excess}(L,z) = \frac{2\pi G(\gamma)r_0^\gamma(L,z)R^{3-\gamma}\phi(L,z)}{3-\gamma}\delta L
\label{equ:sig_exes}
\end{equation}
where $G$ is a constant defined by $\gamma$ (see \citealt{phi85}).

Importantly $\Sigma_{excess}$ is trivial to measure between a sample with redshifts and one without. We may then constrain the clustering strength $r_0(z,L)$ and luminosity function $\phi(z,L)$ of a population with no redshifts.

\section{data}
\label{sec:data}

In this paper we aim to measure the radio luminosity function and clustering strength of high-redshift radio sources. We do this by counting quasar--radio source pairs from a spectroscopic quasar sample that has redshifts and a radio catalogue that has none.

We take the NVSS survey \citep{con98} as our parent sample of radio sources. The NVSS covers the whole sky north of $-30^\circ$, but for our purposes we are only interested in extragalactic sources and so cut out all objects with galactic latitude $|b|<10^\circ$ (as well as $Dec.<-30$). We also make a flux cut at 3\,mJy above which the NVSS is $\sim90$\,\%\ complete \citep{con98} leaving 1,062,117 radio sources in our sample, the vast majority of which have no distance estimate.

The quasar sample we use is a combination of the Sloan Digital Sky Survey (SDSS) DR7 quasar catalogue \citep{sch10b} and the Baryon Oscillation Spectroscopic Survey (BOSS) DR10 quasar catalogue \citep{par13}. We combine the two samples since they cover different redshift ranges: DR7 $0.1<z<2$, DR10 $2<z<3.5$. Neither sample has a single consistent selection function and both are somewhat unevenly distributed across the sky. To flatten the DR7 catalogue we follow the simple cut made by \citep{sch10b} and only allow objects with $i<19.1$: the magnitude limit of the main SDSS quasar survey. For the BOSS sample we make a similar cut of $i<20$ where the number counts begin to turn over. Further cuts to these samples are required to construct accurate random catalogues as discussed in the next section.

%\begin{figure}
%\centerline{\psfig{file=z_dist.ps,width=7.cm,angle=-90}}
%\caption{The redshift distribution of our final quasar sample. The fine lines show the DR7 (black) and DR10 (grey/red) samples separately and the heavy line shows the total. This demonstrates the extra redshift range from $z\sim 2$ to 3.5 made available by the inclusion of the DR10 quasars.}
%\label{fig:zdist}
%\end{figure}

\begin{figure}
\begin{center}
	\includegraphics[width=0.35\textwidth,angle=-90]{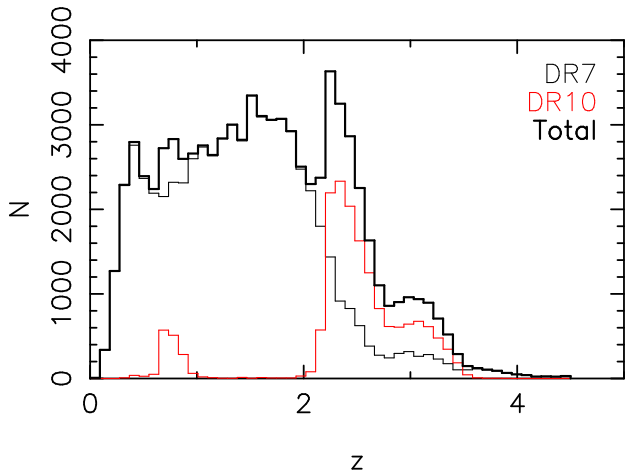}

        \caption{ The redshift distribution of our final quasar sample. The fine lines show the DR7 (black) and DR10 (grey/red) samples separately and the heavy line shows the total. This demonstrates the extra redshift range from $z\sim 2$ to 3.5 made available by the inclusion of the DR10 quasars.}
        \end{center}
        \label{fig:zdist}
\end{figure}

%\begin{figure}
%\centerline{\psfig{file=final_sample_radec.eps,width=9.cm,angle=0}}
%\caption{The distribution of our final quasar samples in the sky. Red and Blue are the SDSS DR7 and BOSS samples respectively while the black line shows the dec.$>-30^\circ$ and galactic $|b|>10^\circ$ cuts that define the NVSS area we consider.}
%\label{fig:sky}
%\end{figure}

\begin{figure}
\begin{center}
	\includegraphics[width=0.45\textwidth]{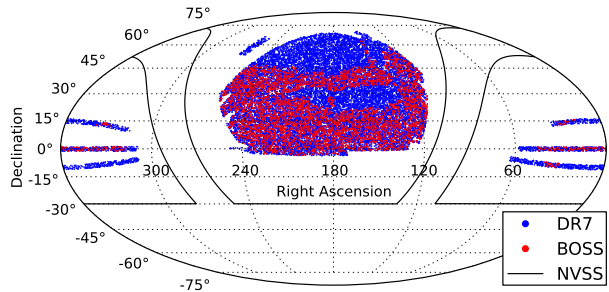}
        \caption{ The distribution of our final quasar samples in the sky. Red and Blue are the SDSS DR7 and BOSS samples respectively while the black line shows the dec.$>-30^\circ$ and galactic $|b|>10^\circ$ cuts that define the NVSS area we consider.}
        \end{center}
       \label{fig:sky}
\end{figure}

\subsection{Random catalogues}

To estimate the excess $\Sigma_{excess}$ we need random catalogues matched to our radio and quasar samples to form a comparison. We create the random radio catalogue by generating random sky positions with $Dec.>-30^\circ$ and $|b|>10^\circ$. We also assign each source a flux drawn from the NVSS at random. In case of any variation in the flux distribution of the NVSS due to the changing beam with declination we discretise the NVSS catalogue into one degree declination strips and only draw a flux value for our random source from objects within the same declination strip.

To create the random quasar catalogue we use {\sc mangle} and the `SDSS DR72' radial selection function from the mangle website \citep{ham04,bla05,swa08}. Note the DR72 mask was developed to reproduce the sky coverage of the main DR7 spectroscopic galaxy survey, and does not include additional fields that were in the DR7 quasar catalogue. Therefore we apply this mask to cut the area of our real DR7 quasar catalogue as well. We then produce a random catalogue from the DR72 mask with ten times the number of random objects as quasars.

To make the random BOSS DR10 sample we again use {\sc mangle} and the same DR72 mask. Note this mask does not include approximately a quarter of the BOSS survey that was only covered photometrically after DR7. However, the DR72 mask reproduces the small-scale coverage of the survey and so we accept this loss of objects.

To cut the DR72 area to just that observed by BOSS we take the field centers of the spectroscopic observations from the SDSS website and only include objects within $1.49^\circ$ of a field center. Again, we produce this random catalogue with ten times the number of objects as BOSS quasars.

Our final quasar catalogue has 80,494 objects, 63,682 from DR7 and 16,812 from DR10. Figure~\ref{fig:zdist} shows the redshift distributions of the final samples split by their survey. Clearly the inclusion of the BOSS DR10 quasars extends the redshift coverage of our sample from $z\sim 2.2$ to 3.5. The distribution of these quasars on the sky along with the NVSS boundaries are shown in Figure~\ref{fig:sky}

\section{Analysis}
\label{sec:anal}

The excess number of radio sources around quasars at a given redshift and radio luminosity (calculated assuming the redshift of the quasar), $\Sigma_{QR}(z,L)$, constrains the cross-clustering strength $r_{0QR}(z,L)$ and the radio luminosity function $\phi(z,L)$ (Equation~\ref{equ:sig_exes}). By assuming prior knowledge of either \rqr\ or $\phi$ we can then constrain the other. In this section we describe models we will assume for \rqr\ and our method for estimating $\Sigma$.

\subsection{The clustering strength of quasars and radio sources}

In reality the cross-clustering strength \rqr\ that appears in Equation~\ref{equ:sig_exes} is rarely measured. More commonly the autocorrelation strengths of quasars, $r_{0QQ}$, or radio sources, $r_{0RR}$, are studied. We will relate these quantities by assuming linear bias such that $r_{0QR}^2 \sim r_{0QQ} \times r_{0RR}$ (e.g. \citealt{wak08b}) and model $r_{0QQ}$ and $r_{0RR}$ as a function of redshift and luminosity based on recent analyses.

Studies of quasar clustering have repeatedly shown that the quasar correlation function is roughly independent of quasar luminosity \citep{daa08,sha11}. $r_{0QQ}$ increases slowly with redshift and, since the mass clustering is falling as redshift increases, the quasar bias rises quickly with redshift. Converting bias into mass via Press-Schechter theory implies that the average dark haloe mass of quasars is roughly constant with $M_{DH}\sim10^{12}$\,M$_\odot$ at all redshifts \citep{croom05,mye06,ros09}.

%\begin{figure}
%\centerline{\psfig{file=Qr0_z_plot1.eps,width=7.cm}}
%\caption{The clustering strengths $(r_0)$ for the populations we are considering. Black points give the quasar autocorrelation strength from \citet{ros09} and the solid line is our simple fit to these. The dashed black line gives the constant radio galaxy autocorrelation strength from \citet{me11}. Grey (red online) lines show the correlations strengths assumed in the SKADS simulation: solid radio-quiet quasars, dashed FRIs and dotted FRIIs.}
%\label{fig:Qr0_z}
%\end{figure}

\begin{figure}
\begin{center}
	\includegraphics[width=0.49\textwidth]{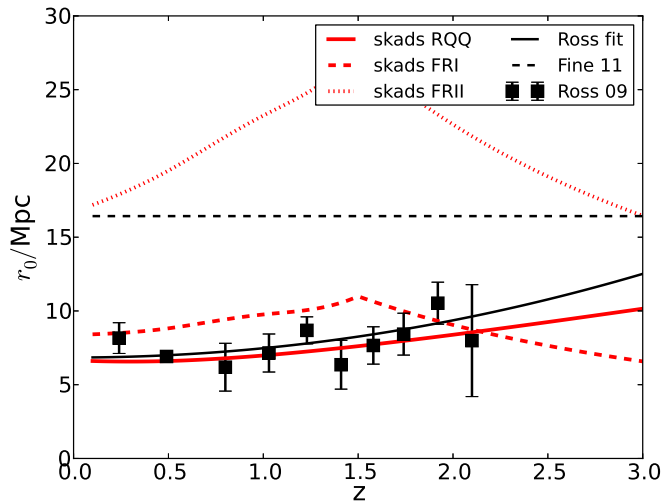}
        \caption{ The clustering strengths $(r_0)$ for the populations we are considering. Black points give the quasar autocorrelation strength from \citet{ros09} and the solid line is our simple fit to these. The dashed black line gives the constant radio galaxy autocorrelation strength from \citet{me11}. Grey (red online) lines show the correlations strengths assumed in the SKADS simulation: solid radio-quiet quasars, dashed FRIs and dotted FRIIs.}
        \end{center}
      \label{fig:Qr0_z}
\end{figure}

Assuming there is no variation in quasar clustering strength with luminosity we only need the variation with redshift. Figure~\ref{fig:Qr0_z} shows the autocorrelation clustering strength of quasars as a function of redshift from \citet{ros09}. We perform a $\chi ^2$ minimisation for evolution of the clustering strength assuming the quadratic form
\begin{equation}
r_0 = a + bz^2, 
\end{equation}
\tcol{where $r_0$ has units of Mpc assuming $h=0.7$}. We  find $a=6.8\pm 0.31$ and $b=0.63 \pm 0.26$. We will use this empirical fit to estimate $r_{0QQ}(z)$.

The clustering of mJy radio sources has been extensively studies at redshifts below $\sim0.8$ with samples cross-matched to optical spectroscopic or photometric galaxies \citep{p+n91,bra05,wak08a,don10,me11,lin14a}. At the radio luminosities sampled in those surveys ($L \gtrsim 10^{24}$\,W/Hz) the radio population is dominated by AGN typically hosted by LRGs. \citet{me11} showed little evolution in the clustering strength of these objects with a constant $r_{0RR}\sim 11.5\,h^{-1}$\,Mpc \tcol{\citep{me11}}. This broadly matches the lack of clustering evolution seen in optically selected LRGs \citep{bel04,wak06,bro07}. At redshifts greater than 0.8 there are few indications of the clustering of radio sources due to the lack of wide-field optical galaxy samples in this redshift regime. \citet{lin14b} used photometric galaxy sample over a small field with deep radio observations to measure the cross correlation with IR galaxies. They found no variation in clustering strength with radio power and while their correlation increased with redshift this is primarily driven by evolution in their IR galaxy sample rather than the radio.

The dependence of clustering strength on radio luminosity is not well described. In their angular correlation analysis of the NVSS \citet{ove03} derived a correlation scale length of $r_{0RR}\sim6\,h^{-1}$\,Mpc for lower luminosity sources $\lesssim 10^{26}$\,W/Hz while the brighter, potentially FRII, sources had a scale length $r_{0RR}\sim14\,h^{-1}$\,Mpc. On the other hand clustering analyses of radio surveys matched to optical galaxies have found no luminosity dependence in the large-scale haloe-haloe regime \citep{don10,me11,lin14b}. Note that since the \citet{ove03} had no redshift information a considerable series of assumptions about the radio population were required to derive their result, on the other hand both \citet{don10} and \citet{me11} struggled for sources in their samples with $L>10^{26}$\,W/Hz while \citet{lin14b} had none.

Given the few constraints on $r_{0RR}$ we initially make the simplest empirical assumption. That is the clustering strength of our radio sample is constant with redshift and luminosity with $r_{0RR}=11.5\,h^{-1}$\,Mpc. Assuming linear bias and $r_{0QQ}(z)$ from \citet{ros09} this makes up our empirical (EMP) model for $r_{0QR}(L,z)$.

As an alternative and check we also consider the values assumed in \citet{wilm08} when they were attempting to model the radio sky. They followed \citet{ove03} and assumed considerably stronger clustering for the brightest radio sources. For $z<1.5$ they assumed constant dark haloe masses of $10^{13}$ and $10^{14}h^{-1}$\,M$_\odot$ for FRI and FRII sources respectively. At high redshift the clustering strength of their FRII sources would become unphysically large and so for $z>1.5$ they held the bias of their FRI and FRII sources constant. We make the simplistic assumption that all radio sources with $L<10^{26}$\,W/Hz are FRI sources, the rest being FRIIs. For radio-quiet AGN, essentially quasars, they assumed a constant haloe mass of $3\times 10^{12}h^{-1}$M$_\odot$ with a similar redshift cut at $z=3$ above which the bias was held constant. We will refer to this alternative model for $r_{0QR}$ as the W08 model (see Figure~\ref{fig:Qr0_z} for a comparison of the differing models).

%\begin{figure}
%\centerline{\psfig{file=nice_angcor1.ps,width=9.5cm,angle=-90}}
%\caption{The quasar--radio source angular correlation function of our sample. The vertical lines are at 2 and 20\,Mpc projected distance. The redshift limits for each bin are given in the bottom left of the panels.}
%\label{fig:angcor}
%\end{figure}

\begin{figure}
\begin{center}
	\includegraphics[width=0.37\textwidth,angle=-90]{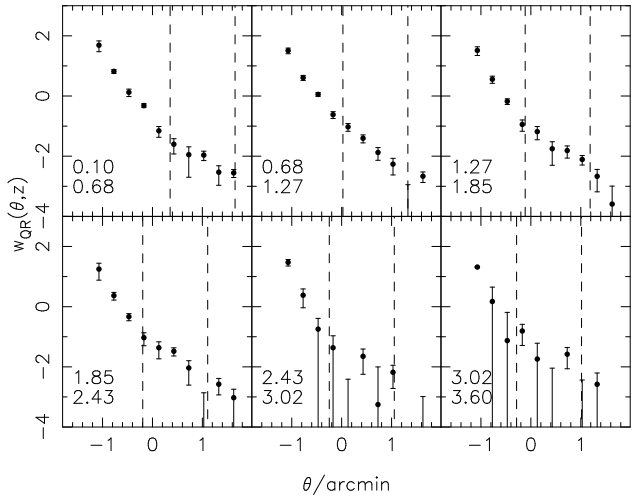}
        \caption{ The quasar--radio source angular correlation function of our sample. The vertical lines are at 2 and 20\,Mpc projected distance. The redshift limits for each bin are given in the bottom left of the panels.}
        \end{center}
     \label{fig:angcor}
\end{figure}

\subsection{Removing radio-loud quasars}

The statistic $\Sigma_{excess}$ defined in section~\ref{sec:oot} is the excess number of radio sources around quasars. In the derivation of Equation~\ref{equ:sig_exes} it is assumed that this excess comes only from the clustering of matter. Radio-loud quasars in our sample increase the measured $\Sigma$ and bias our results. To illustrate this effect Figure~\ref{fig:angcor} shows the angular cross correlation function $w_{QR}(\theta,z)$ for our sample of quasars (split into redshift bins) and the NVSS catalogue. The vertical lines in the figure show 2 and 20\,Mpc projected onto the angular scale at the redshift of the bins. The upturn below $\sim 2$\,Mpc is caused by a combination of radio-loud quasars and non-linear single haloe clustering (e.g. \citealt{b+w02}). We remove the radio-loud contribution by only counting pairs in the annulus between $R=2$ and 20\,Mpc. In this region the angular correlation function is well fit by a single power law indicating that the spatial correlation function $\xi(r)$ is also approximately a power law. We choose 2\,Mpc as the lower limit both from inspection of Figure~\ref{fig:angcor} and since this corresponds to roughly the largest known giant radio galaxies \citep{sar05}.

\subsection{Calculating $\Sigma$}

To estimate $\Sigma$ we count all radio sources with a projected distance between 2 and 20\,Mpc from a quasar in our samples, $N_{DD}$. In addition to these data-data pairs we also substitute our random catalogues and count $N_{DR}$, $N_{RD}$ and $N_{RR}$. This is done in redshift and luminosity bins where the luminosity of the radio sources are calculated assuming the redshift of the quasar. Redshift bins are equally spaced over the interval sampled by our quasars $0.1<z<3.6$. Luminosity bins are logarithmically spaced over three orders of magnitude, the lower limit of which is the lowest luminosity observable in that redshift bin. The flux limit of the NVSS catalogue gives a Malmquist bias. Hence in our summations each pair is weighted by $V_{bin}/V_{max}$.

Following \citet{ham93} we estimate $\Sigma$ with
\begin{equation}
\Sigma = \frac{1}{N_Q}(N_{DD} - N_{RR}N_{DR}/N_{RD}),
\end{equation}
where $N_Q$ is the total number of quasars in the redshift bin. From Equation~\ref{equ:sig_exes} we relate $\Sigma$ to the luminosity function and clustering strength with
\begin{equation}
\Sigma_{excess}(L,z) = \frac{2\pi G(\gamma)r_0^\gamma(L,z)(R_{max}^{3-\gamma}-R_{min}^{3-\gamma})\phi(L,z)}{3-\gamma} \delta L.
\label{equ:ourS1}
\end{equation}
Throughout this paper we will assume $\gamma=1.8$ hence $G=3.678$, and use $R_{max},R_{min}=20,2$\,Mpc. Equation~\ref{equ:ourS1} becomes
\begin{equation}
\Sigma_{excess}(L,z)  =  657 r_{0QR}^{1.8}(L,z)\phi(L,z) \delta L.
\label{equ_sig_sim}
\end{equation}
We assume this simple relationship between $\Sigma$, $r_0$ and $\phi$ throughout the rest of this work.

We use jack knife resampling to estimate our errors by splitting our sample into 20 even sized (by number of quasars) sub fields by right ascension. We then calculate $\Sigma$ in each sub field and estimate the `field-to-field' errors from the rms of these values for $\Sigma$ (e.g. \citealt{saw11,me11}).

\section{Excess pair counts}
\label{sec:results}

%\begin{figure}
%\centerline{\psfig{file=nice_sigma.ps,width=8.cm,angle=-0}}
%\caption{The average excess of radio sources around quasars, $\Sigma$, binned by redshift and luminosity. We show the plot with both a linear and logarithmic scale since several of the points are scattered below $\Sigma=0$ by noise.}
%\label{fig:sigma}
%\end{figure}

\begin{figure}
\begin{center}
	\includegraphics[width=0.47\textwidth]{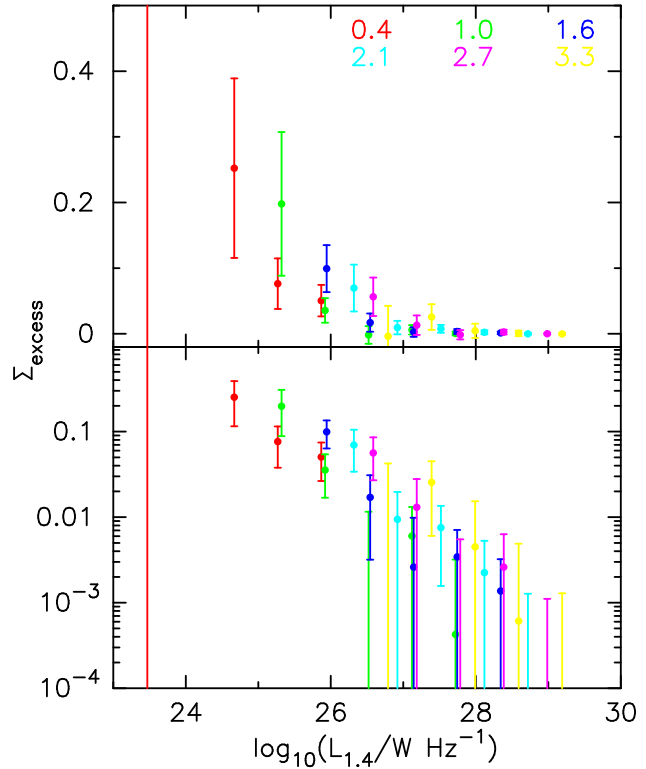}
        \caption{ The average excess of radio sources around quasars, $\Sigma$, binned by redshift and luminosity. We show the plot with both a linear and logarithmic scale since several of the points are scattered below $\Sigma=0$ by noise.}
        \end{center}
     \label{fig:sigma}
\end{figure}

Figure~\ref{fig:sigma} shows the values of $\Sigma_{excess}$ we calculate from our sample for six redshift and five luminosity bins. The way $\Sigma$ is calculated means that it can be scattered to negative values due to noise. Hence we show both a linear and logarithmic scale to illustrate how the measured values and their errors behave. The points with $\Sigma<0$ and their errors still contain information about our sample and need to be included in any analysis to avoid introducing bias. Furthermore the error bars are symmetric and approximately Gaussian in linear space. Hence, while plots may be in log space, any fitting to the data is performed in linear space.

It is apparent from Figure~\ref{fig:sigma} that at fixed radio luminosity $\Sigma(L,z)$ increases slightly with redshift. This indicates that one of $r_{0QR}$ or $\phi$ is increasing with redshift.

%\begin{figure*}
%\centerline{\psfig{file=nice_LF1.ps,width=17.cm,angle=-90}}
%\caption{The radio luminosity function. Solid points assume the EMP clustering model while open points are W08. The solid line shows the \citet{wil01} RLF model and the dashed line shows our evolving power law model fit to the data at the midpoints of the bin (top-right in each panel). The redshift range for each bin is given at the top-right of each panel. The poor fit in the first redshift bin is due to the data being dominated by radio sources at the high-$z$ limit of the bin.}
%\label{fig:LF1}
%\end{figure*}

\begin{figure*}
\begin{center}
	\includegraphics[width=0.7\textwidth,angle=-90]{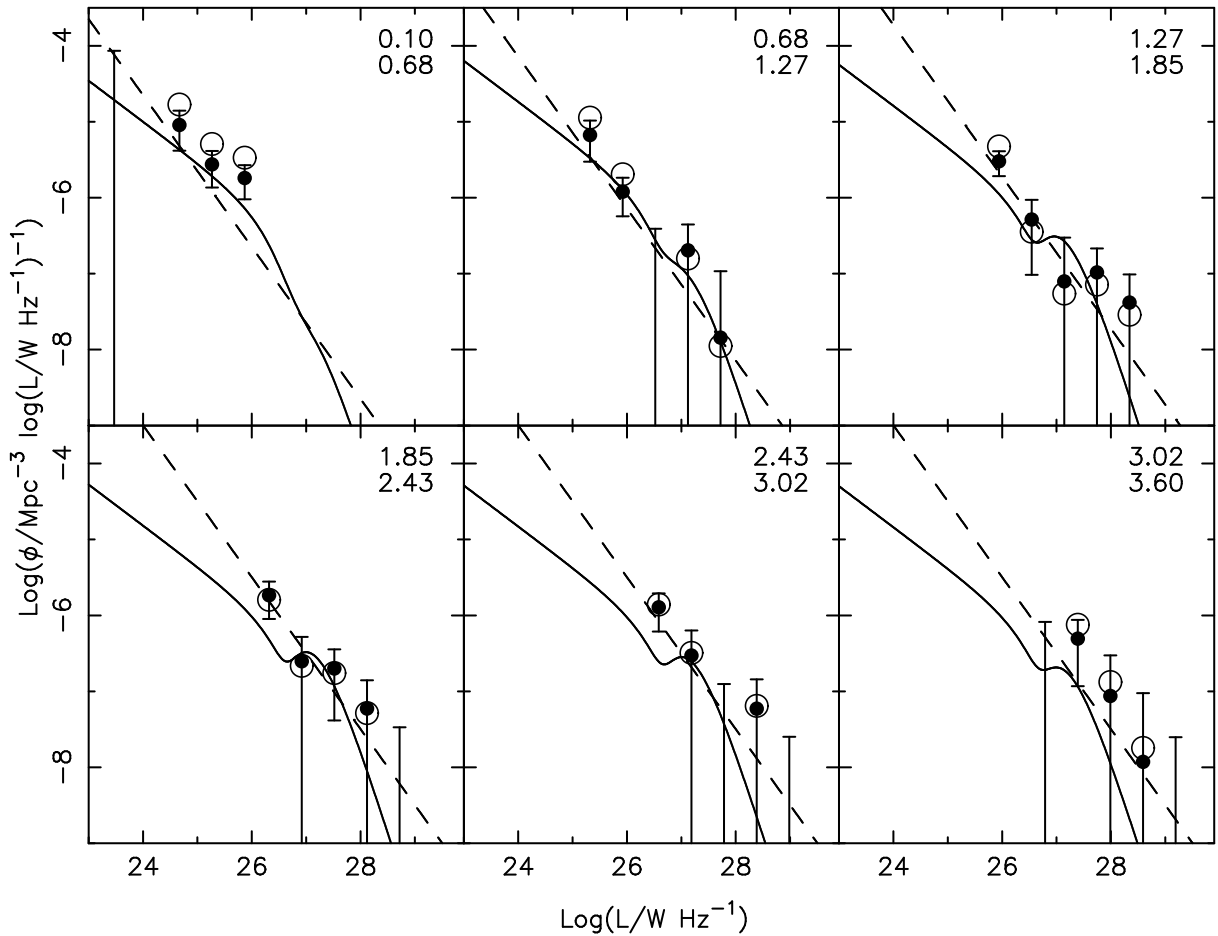}
        \caption{ The radio luminosity function. Solid points assume the EMP clustering model while open points are W08. The solid line shows the \citet{wil01} RLF model and the dashed line shows our evolving power law model fit to the data at the midpoints of the bin (top-right in each panel). The redshift range for each bin is given at the top-right of each panel. The poor fit in the first redshift bin is due to the data being dominated by radio sources at the high-$z$ limit of the bin.}
        \end{center}
    \label{fig:LF1}
\end{figure*}

\section{The radio luminosity function}
\label{sec:rlf}

Figure~\ref{fig:LF1} shows the radio luminosity function we calculate for six redshift bins. We show the LF calculated assuming the EMP (solid points) and W08 (open points) clustering models. There is very little overall difference in the LFs between the clustering models. The solid lines in Figure~\ref{fig:LF1} show the \citet{wil01} RLF model at the mid-point redshift of the bin and it is clear that in general our results are consistent with their model.

To describe our data further we initially fit an evolving power law $\phi = A (1+z)^\alpha(L/10^{26})^\beta$ and find $\alpha = 1.00 \pm 0.35$ for the EMP model and $1.02 \pm 1.05$ for W08. However, we find  the large redshift and flux range we sample mean this is not an accurate model for our data. To better illustrate the redshift evolution in our data we fix the luminosity exponent to that from our fit $(\beta = -0.99 $EMP; -0.85 W08) and just fit for the amplitude of the power law in each redshift bin. Figure~\ref{fig:LF2} shows the amplitude of the fitted power law at $10^{26}$\,W/Hz as a function of redshift.

\subsection{Redshift cutoff}

\begin{figure}
\begin{center}
	\includegraphics[width=0.38\textwidth,angle=-90]{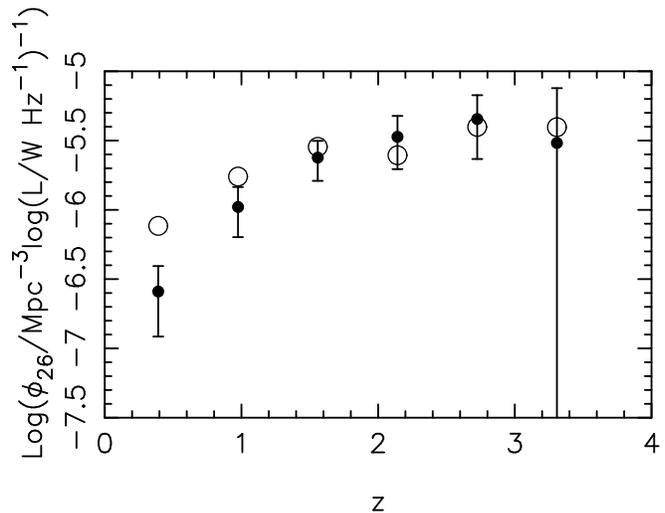}
        \caption{ The evolution of the amplitude of the RLF from $z\sim 3.5$. The solid points are using the EMP model, the open points W08.}
        \end{center}
   \label{fig:LF2}
\end{figure}

\begin{figure}
\begin{center}
	\includegraphics[width=0.49\textwidth]{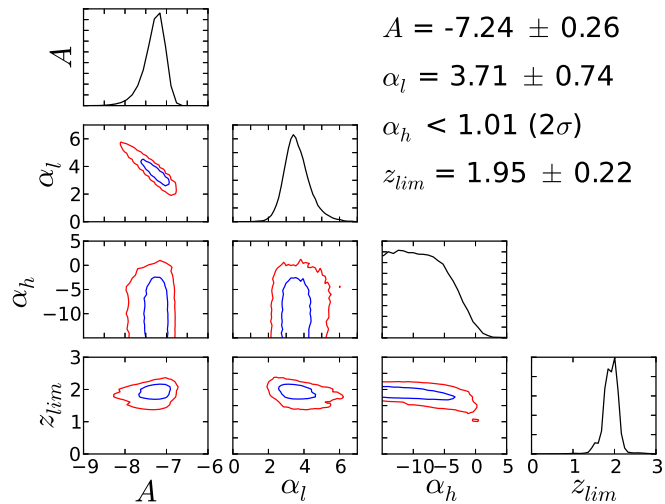}
        \caption{ The results of our MCMC fitting of our RLF data. Blue and red lines show the 1 and 2$\sigma$ marginalised constrains respectively. We constrain all but the high redshift evolution parameter for which we can only obtain an upper limit..}
        \end{center}
   \label{fig:mcmc1}
\end{figure}

Figure~\ref{fig:LF2} indicates that the increase in space density slows and may turnover at higher redshifts. To include this behaviour we introduce a \tcol{redshift limit and separate parameter for high-redshift evolution}
\begin{equation}
\Phi(L,z) =  \bigg\{ \begin{array}{ll}
(1+z)^{\alpha_l}(L/10^{26})^\beta &  z\leq z_{\rm lim} \\
(1+z_{\rm lim})^{\alpha_l-\alpha_h}(1+z)^{\alpha_h}(L/10^{26})^\beta & z>z_{\rm lim}. \\
\end{array}
\label{equ:pl_ev}
\end{equation}
Where $\alpha_h$ and $\alpha_l$ are the evolution parameters above and below $z_{\rm lim}$. Since there can be relatively rapid evolution we bin our data into 25 redshift and 10 luminosity bins. We fit our model with simple Markov chain Monte Carlo (MCMC) routine iterated 500,000 times to find the preferred values. We fix $\beta= -1$ in the fitting since there can be a degeneracy between the evolution parameter and $\beta$ due to our LFs being defined in different parts of luminosity space at different redshifts. Figure~\ref{fig:mcmc1} shows the probability distributions from our fitting using the EMP model, along with the best fit values. Clearly our data support strong $\alpha_h \sim 4$ evolution up to $z_{\rm lim}\sim 2$. At high redshift we only have an upper limit on the evolution parameter but can show at least that the increase in space density stops, or turns over. Interestingly this redshift cut-off is almost identical to that found by \citet{wil01} for their high-luminosity objects $z_{\rm cut-off}=1.91\pm0.16$.

%\begin{figure}
%\centerline{\psfig{file=phi26_z.ps,width=7.cm,angle=-90}}
%\caption{The evolution of the amplitude of the RLF from $z\sim 3.5$. The solid points are using the EMP model, the open points W08.}
%\label{fig:LF2}
%\end{figure}
%
%\begin{figure}
%\centerline{\psfig{file=evPlM1cut_1.eps,width=10.cm,angle=0}}
%\caption{The results of our MCMC fitting of our RLF data. Blue and red lines show the 1 and 2$\sigma$ marginalised constrains respectively. We constrain all but the high redshift evolution parameter for which we can only obtain an upper limit.}
%\label{fig:mcmc1}
%\end{figure}

To compare further with \citet{wil01} we fitted out data using a parametrisation based on their models. Their model `C' is split into to population roughly separated at $L=26$\,W/Hz. We fit for $(1+z)^\alpha$ evolution with a redshift cut as in Equation~\ref{equ:pl_ev} for each population. In this fitting we convert from their cosmology to our own to make the fitted values comparable. In the high luminosity regime we have very few quasar-radio source pairs and consequently this part of parameter space is poorly constrained. On the other hand at low luminosities we find  $(1+z)^{2.35\pm0.89}$ evolution to redshift $1.94\pm0.43$ above which we can only estimate an upper limit for the evolution parameter $\alpha_h<1.3$ (2$\sigma$). This contrasts with their findings of $\alpha_l=3.5$ up to a redshift cutoff at 0.7. \citet{wil01} have a considerably brighter sample than we use here and their redshift cut is imposed by their flux limit. It may be that since we are able to better define $z_{cut}$ this explains the smaller discrepancy between our evolution parameters.

\subsection{Discussion}
%\label{sec:disc}

Our results are consistent with a model that evolves strongly, $\Phi\propto(1+z)^{3.7}$, to $z\sim 2$ above which the LF either stays constant or falls. Interestingly, this is approximately the same redshift evolution that \citet{wil01} found for their high-luminosity population of sources. The indication is that, rather than having two separate populations with a \tcol{transition} at $L\sim10^{26}$\,W, the radio LF may be dominated by a single population in the luminosity-redshift regime we are sampling.

Recent studies of the RLF have focused on the accretion mechanisms that launch the radio jet and the role that the AGN may play in heating the intergalactic medium. Terminologies differ but we will refer to high-excitation radio galaxies (HERGs) associated with high-accretion rate optical AGN and low-excitation radio galaxies (LERGs) associated with substantially lower accretion rates via advection-dominated accretion flows in massive elliptical galaxies. At low redshifts LERGs dominate the LF below $L\sim10^{26}$W \citep{hec07,b+h12}. At these fainter luminosities and lower redshifts the LF has been shown to only evolve slowly \citep{c+j04,sad07,smo09,mca13} However, there is evidence that the HERG population evolves considerably more strongly than the LERGs \citep{wil01,bes14}, potentially becoming the dominant population in the luminosity range we sample around $z\sim1$.

The 3\,mJy flux limit we impose allows us to sample luminosities of $10^{26}$\,W up to $z=2.5$. However, the strong evolution of the LF coupled with the increased comoving volume at high redshift means that we are dominated by sources at $z\sim2$. Assuming the simplistic power law LF from our MCMC fit, less than $5$\,\%\ of our sample is at $z<1$. The indication is that our signal is dominated by HERGs, and hence it may be unsurprising that we find almost identical evolution parameters to the factor of $\sim50$ brighter \citet{wil01} sample.

\citet{wad01} and \citet{rig11} found that the fainter end of the RLF peaked in density at lower redshift. Due to the nature of our analysis we are always dominated by the radio sources close to our flux limit. Hence we cannot split the sample into luminosity bins to compare across a range of redshifts. We find a redshift cutoff at $z=1.95\pm0.22$. At this redshift our 3\,mJy flux limit translates to $\log(L/{\rm W})=25.77$, and so we can consider the turnover seen in our data to be due to radio sources at or somewhat brighter than this. At these luminosities \citet{rig11} found a redshift cut closer to $z=1$ although our results are consistent within a few sigma.

\section{The clustering of radio sources}
\label{sec:clust}

%\begin{figure}
%\centerline{\psfig{file=nice_r0.ps,width=7.cm,angle=-90}}
%\caption{The points with error bars show $r_{0QR}$ as a function of redshift for our sample. The dashed black line gives the implied value of $r_{0RR}$ assuming the empirical values of $r_{0QQ}$ from \citep{ros09}. Grey (red online) dashed and dotted lines show our EMP model values for $r_{0QQ}$ and $r_{0RR}$ respectively.}
%\label{fig:Rr0_z}
%\end{figure}
%

\begin{figure}
\begin{center}
	\includegraphics[width=0.34\textwidth,angle=-90]{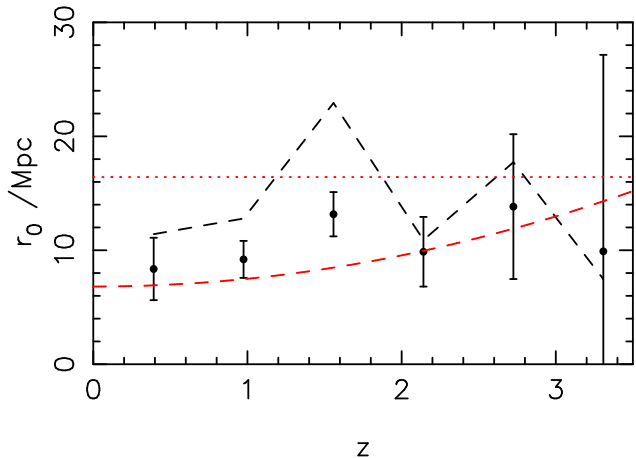}
        \caption{ The points with error bars show $r_{0QR}$ as a function of redshift for our sample. The dashed black line gives the implied value of $r_{0RR}$ assuming the empirical values of $r_{0QQ}$ from \citep{ros09}. Grey (red online) dashed and dotted lines show our EMP model values for $r_{0QQ}$ and $r_{0RR}$ respectively.}
        \end{center}
   \label{fig:Rr0_z}
\end{figure}

Reversing what we have done above we can integrate the \citet{wil01} luminosity function above our flux limit to give $\phi$ in Equation~\ref{equ_sig_sim} and hence constrain the clustering strength between the radio sources and quasars in our sample. Figure~\ref{fig:Rr0_z} shows the measured cross correlation strength in six redshift bins. While there may be some hint of an increase in clustering strength with $z$, $r_{0QR}$ is consistent with a constant value of $10.4\pm 2.5$\,Mpc over the full redshift range sampled (if poorly constrained at the highest redshifts). The dashed line in Figure~\ref{fig:Rr0_z} shows the value of $r_{0RR}$ assuming the empirical fit in Figure~\ref{fig:Qr0_z} and $r_{0QR}^2=r_{0QQ}r_{0RR}$. Again the estimates for $r_{0RR}$ are consistent with a constant value of $15.4$.

At lower redshift ($z<1.5$) we find considerably stronger clustering in our sample compared to quasars, more in line with the results for radio galaxies from \citet{me11} and \citet{lin14a}. At higher redshifts both our errors and the clustering strength of quasars increase and we cannot form a distinction.

Our results for $r_{0RR}$ are not consistent with the strong $20-25$\,Mpc values assumed in \citet{wilm08} for FRII sources. This is despite our being dominated by bright $L>10^{26}$\,W sources at high redshift and our sample potentially being dominated by HERGs/FRIIs at all redshift as discussed in our RLF analysis.
None the less, we find are strong enough clustering to indicate these radio sources are in some of the most massive haloes at all redshifts we sample ($\sim 10^{14}$\,M$_\odot$ at $z\sim0$ to $\sim 10^{12.5}$\,M$_\odot$ at $z\sim3$). A possible explanation for this would be a later ($z<1.5$) break imposed in the \citet{wilm08} bias/clustering model, combined with our being dominated by fainter FRI/LERG sources at low redshift. Alternatively the clustering strength could depend strongly on luminosity and redshift to contrive to give our results, although this has not been noted before.

\section{Summary}
\label{sec:sum}

\tcol{We measure the overdensity $\Sigma(z,L)$ of radio sources around spectroscpic quasars and relate this to evolution in the radio source population from $z\sim3.5$ to today. Our key results can be summarised:}

{\bf  $\Sigma(z,L)$ is measured in redshift/luminosity bins and we find significant evolution with redshift.}
This can only be explained by either the RLF of clustering strength increasing to $z\sim2$.

{\bf Under some simple models for $r_0(z,L)$ we find strong evolution, $\phi\propto (1+z)^{3.7\pm0.7}$, up to $z=1.9\pm0.2$ above which the evolution declines, although we can only constrain an upper limit.}
These evolution parameters are consistent with those found by \citet{wil01} for the brighter radio source population. The indication may be that the same population of HERGs dominates the NVSS at all flux densities above $z\sim1$.

{\bf Assuming the \citet{wil01} LF model we find the clustering strength of radio sources to be consistent with a value of $r_{0RR}=15.0 \pm 2.5$}
This is inconsistent with quasars at low redshift and the W08 model for FRII clustering at intermediate ($1\lesssim z \lesssim 2$). A possible explanation would be the population being dominated by LERGs at low redshift and clustering more like quasars at higher redshift. Regardless, our results show that these radio sources are found in the most massive dark matter haloes at all redshift we sample.

In this work we have demonstrated a technique that exploits a well defined sample with distance information to constrain the luminosity function and clustering of a sample without. The next generation of radio surveys will push still deeper beyond the flux limits of the NVSS used here. Despite new wide-field redshift surveys (e.g. EUCLID) the vast majority of the sources detected in these surveys will not have reliable distances. The method presented in this work offers an alternative approach to studying these populations with observational strategies that are already possible and, for the most part, have already been carried out.

\section{Acknowledgments}

SF and RJ would like to acknowledge SKA South Africa and the NRF for their funding support.

\bibliography{bibRJ}

\end{document}